\documentclass[a4paper,twoside,openbib,10pt]{article}
\usepackage{graphicx}
\usepackage{float}
\usepackage{layout}

\begin{document}
\renewcommand{\thefootnote}{\fnsymbol{footnote}}
\setlength{\bibindent}{0.25in}
\renewcommand{\figurename}{\small \textbf{Figure}}

\vspace*{96pt}

{\flushright \Large\bf COUPLING AND DEPHASING IN JOSEPHSON
CHARGE-PHASE QUBIT WITH RADIO FREQUENCY READOUT} \vspace{24pt}
{\large \flushleft Alexander B. Zorin \footnote{A. B. Zorin,
Physikalisch-Technische Bundesanstalt, \protect\hbox{38116}
Braunschweig, Germany. }} \vspace{24pt}

\thispagestyle{plain}

 \begin{abstract}
The Cooper pair box qubit of a two-junction-SQUID configuration
enables the readout of the qubit states by probing the effective
Josephson inductance of the SQUID. This is realized by coupling
the qubit to a high-Q tank circuit which induces a small
alternating supercurrent in the SQUID loop. The effect of a small
(but finite) geometrical inductance of the loop on the eigenstates
of the system is figured out. The effect of qubit dephasing due to
quadratic coupling to the tank circuit is evaluated. It is shown
that the rate of dephasing in the vicinity of the magic points is
relatively low unless the Josephson junctions forming the qubit
are rather dissimilar. In the vicinity of the avoided
level-crossing point such dephasing is always significant.
 \end{abstract}


\section{Introduction}

\pagestyle{myheadings} \markboth{\rm A. B. Zorin}{\bf COUPLING AND
DEPHASING IN JOSEPHSON \dots}

The readout device is a critical component of any potential
quantum computing circuit. For the Josephson qubits there is a
number of sensitive cryogenic devices available (SQUIDs, switching
Josephson junctions, single electron transistors and traps, etc.)
enabling the readout of the qubit state. However, operation of
these devices is usually associated with a significant exchange of
energy between detector and qubit, so in order to avoid fast
decoherence the detector must be reliably decoupled from the qubit
at the time of quantum manipulation. Recently, the class of
Josephson qubit detectors based on the measurement of the reactive
component of electrical signals related to nonlinear behavior of
the Josephson inductance has been extensively studied
\cite{Zorin-PhysC, Ilichev-FNT, Lupascu, Siddiqi}. Due to specific
coupling to the qubit variables and non-dissipative
characteristics of the Josephson supercurrent, these circuits can
have a much weaker backaction and, therefore, cause lesser
decoherence. Moreover, these circuits may possibly enable quantum
nondemolition measurements of a Josephson qubit \cite{Averin-QND}.

\begin{figure}
\begin{center}
\includegraphics[width=9cm]{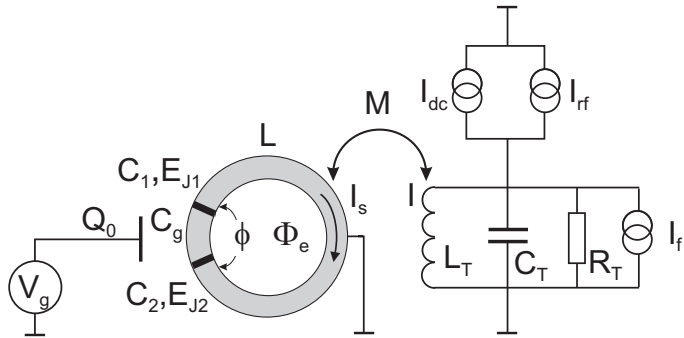}
\end{center}
\caption{ Electric circuit diagram of the charge-phase qubit. The
core of the qubit is the superconducting ring of small inductance
$L$ including two small tunnel junctions of capacitances $C_1$ and
$C_2$ and Josephson coupling energies $E_{J1}$ and $E_{J2}$,
respectively. The small island between the junctions is polarized
by charge on small gate capacitance $C_g$. The ring is inductively
(mutual inductance $M$) coupled to an rf-driven parallel tank
circuit. Voltage across the tank circuit is read out by a
(presumably cold) preamplifier which is not shown.} \label{Scheme}
\end{figure}

In this paper, we consider the charge-phase qubit
\cite{Zorin-PhysC} comprising a macroscopic superconducting ring
including two small Josephson junctions with a small island in
between (see Fig.~1), i.e., a Cooper pair box \cite{Bouchiat} of
SQUID configuration. This setup is, in principle, similar to that
of Quantronium \cite{Vion}, in which, however, a larger additional
Josephson junction was inserted in the superconducting loop for
the purpose of readout by switching this junction into a
finite-voltage state. In our case, the ring of inductance $L$ is
inductively coupled to the tank circuit that enables both phase
control and readout of the qubit. The special convenience of this
simple setup is that it can function even without uncoupling the
tank circuit from the qubit. It has recently been shown that
operation of this qubit in the so-called magic points made it
possible to decouple (in linear approximation) both the diagonal
and the off-diagonal components of the variables of the measuring
system (environment) from the qubit \cite{Zorin-JETP}. Moreover,
the resonance shape of the noise spectrum allowed significant
reducing relaxation and, in principle, performing single shot
measurements. Here we analyze the quadratic effect of this noise
and show that dephasing caused by this effect in the vicinity of
the magic operation points is small, provided the qubit is
sufficiently symmetric.

\section{Qubit parameters and the model}

For the vanishingly small value of inductance $L$ the behavior of
the charge-phase qubit is primarily determined by the relationship
between the energies associated with the island and the junctions.
These are the Coulomb energy of the island, $E_c=e^2/2C_{\Sigma}$,
and the Josephson coupling energies of the junctions, $E_{J1,J2}=
\hbar I_{c1, c2}/2e$, where $C_{\Sigma}=C_1+C_2+C_g$ is the
island's total capacitance including the capacitances of the
individual junctions $C_1$ and $C_2$ and the gate capacitance $C_g
\ll C_{1,2}$; $I_{c1, c2}$ are the nominal values of the Josephson
critical currents. We assume that
\begin{equation} \label{EJEckT} (\Phi_0/2\pi)I_{c0} =
E_{J0} \equiv (E_{J1}+E_{J2})/2 \simeq E_c,
\end{equation}
where $\Phi_0=h/2e$ is the flux quantum, $E_{J0}$ and $I_{c0}
=(I_{c1}+I_{c2})/2$ are the average Josephson coupling energy and
the average critical current, respectively. The coupling energies
of individual junctions are almost similar, $E_{J1}\approx
E_{J2}$, so the values of the dimensionless parameters $j_{1,2}
\equiv E_{J1,J2}/2E_{J0} \approx 0.5$, while $j_1+j_2=1$. The
junction capacitances are also assumed to be almost similar,
$C_{1} \approx C_{2}$, and the dimensionless parameters
$\kappa_{1,2} \equiv C_{2,1}/C_{\Sigma} \approx 0.5$, while
$\kappa_1+\kappa_2=1$. The qubit island is capacitively coupled to
the gate voltage source $V_g$ which controls the polarization
charge $Q_0 = C_g V_g$. The inductively coupled coil $L_T$
carrying current $I$ induces the external magnetic flux
$\Phi_e\equiv (\Phi_0/2\pi)\phi_e =-M I$ applied to the qubit loop
and produces the sum phase bias $\phi \equiv \varphi_1+\varphi_2 =
\phi_e$. The core Hamiltonian includes the charging (kinetic)
energy and Josephson (periodic potential) term,
\begin{equation} \label{H0} H_0= 4E_c (n-Q_0/2e)^2 - E_J(\phi)\cos \chi,
\end{equation}
where the operator $n$ of the number of excess Cooper pairs on the
island is
\begin{equation} \label{n}
n = -i \frac{\partial}{\partial \varphi} = -i
\frac{\partial}{\partial \chi},\qquad [n,\varphi]=[n,\chi]=i.
\end{equation}
The operators of the island phases are
$\varphi=(\varphi_1-\varphi_2)/2$ and $\chi=\varphi+\gamma(\phi)$,
where the angle deviation due to asymmetry of the junctions
$\gamma = \arctan \left[(j_1-j_2) \tan(\phi/2) \right]$. The
amplitude of the effective Josephson coupling energy of two
junctions is equal to
\begin{equation} \label{EJ} E_J(\phi) =
\left( {E_{J1}^2 + E_{J2}^2 + 2E_{J1} E_{J2} \cos \phi }
\right)^{1/2}.
\end{equation}

The two lowest eigenvalues of energy, $E_0(Q_0,\phi)$ (ground
state) and $E_1(Q_0,\phi)$ (first excited state), with
corresponding eigenstates $|0\rangle$ and $|1\rangle$, form the
basis suitable for qubit operation. In this basis the Hamiltonian
Eq.\,(\ref{H0}) takes the diagonal form
\begin{equation} \label{epsilon} H_0 = - (\epsilon/2) \sigma_z,
\end{equation}
where energy $\epsilon(Q_0,\phi) \equiv \hbar \Omega = E_1-E_0$
and $\sigma_z$ is the Pauli matrix. The energy landscape is $2e$
periodic in $Q_0$ and $2\pi$ periodic in $\phi$. The plots can be
found elsewhere (see, for example, Refs.\,\cite{Vion} and
\cite{Zorin-JETP}). The reverse Josephson inductance of the qubit
in the ground (excited) state $L_{0,1}^{-1}$ is determined by the
local curvature of the corresponding energy surface,
\begin{equation} \label{L_J} L_{0,1}^{-1}(Q_0,\phi) =
\left(\frac{2\pi}{\Phi_0} \right)^2 \frac{\partial^2
E_{0,1}(Q_0,\phi)}{\partial \phi^2}.
\end{equation}
Depending on the value of the ratio $E_{J0}/E_c$ and the
relationship between the Josephson energies of the junctions,
$j_1/j_2$, the inductance $L_{0,1}$ can take either positive or
negative values in different points on the $Q_0$-$\phi$ plane. In
the so-called magic points (extremum or saddle points), i.e.,
$A\;(Q_0=0,\:\phi=\pi)$, $B\;(Q_0=0,\: \phi=0)$ and $C\;(Q_0=e,\:
\phi=0)$, the absolute values of $L_{0,1}^{-1}$ achieve local
maxima, while in the avoided level-crossing point $D\;(Q_0=e,\:
\phi=\pi)$ its value is the largest \cite{Zorin-JETP}.
Specifically, in point $D$, where effective coupling is small,
$E_J(\pi)=|E_{J1}-E_{J2}| \ll E_c$, these values for two states
are equal to \cite{Krech}
\begin{equation} \label{L01-D}
L_{0,1}^{-1}(D)=\mp \frac{2\pi}{\Phi_0}\frac{j_1
j_2}{|j_1-j_2|}I_{c0}.
\end{equation}

The drive frequency $\omega_{\rm rf}$ (close to the resonance
frequency of the tank circuit $\omega_T = (L_T C_T)^{1/2} \ll
k_BT/\hbar$) is much lower than the qubit frequency $\Omega$, so
the induced classical oscillations of phase $\phi$ are adiabatic.
Due to coupling to the ring, the effective inductance seen by the
tank circuit is \cite{Rifk}
\begin{equation} \label{L-eff} L_{\rm eff}^{(0,1)}=
L_T - \frac{M^2}{L+L_{0,1}} \approx L_T - \frac{M^2}{L_{0,1}}.
\end{equation}
As a result, the resonance frequencies take the distinct values
for the ground and excited states, $\omega_{0,1}=(L_{\rm
eff}^{(0,1)}C_T)^{-1/2}$, and this property is used for the radio
frequency readout of the qubit \cite{Zorin-PhysC}.

\section{Effect of finite inductance of the ring}

The problem of the inductance effects in the persistent current
flux qubits had been addressed by Crankshaw and Orlando in
Ref.\,\cite{Crankshaw} where the corrections to the energy level
values of these qubits were found. Recently, Maassen van den Brink
had analyzed the finite inductance effect in the
three-Josephson-junction jubits considering the self-flux as a
"fast variable" \cite{MaassenVanDeBrink}. He had shown that for
small inductances this effect is merely reduced to renormalization
of the individual Josephson coupling energies entering the system
Hamiltonian.

 In the case of the charge-phase qubit, the sum phase
$\phi$ is also no longer a good variable. Instead, its combination
with the normalized circulating current (which is a
quantum-mechanical operator) is controlled by the external flux
$\Phi_e$,
\begin{equation} \label{LIs} \phi
+\frac{2\pi}{\Phi_0}LI_s(\phi,\chi)= \phi_e.
\end{equation}
The operator of the circulating current can be presented as
\cite{Zorin-JETP}
\begin{equation} \label{IsMod} I_s=
I_\parallel(\phi) \cos\chi +I_\perp(\phi) \sin\chi.
\end{equation}
In the given basis, the operator $\cos\chi$ is diagonal and
$\sin\chi$ is off-diagonal, so the amplitudes of the longitudinal
$(I_\parallel)$ and transversal $(I_\perp)$ components of current
$I_s$ are
\begin{equation}
 \label{I-1}
I_\parallel= \frac{8\pi}{\Phi_0}\, j_1 j_2\sin\phi\,
\frac{E_{J0}^2}{E_J(\phi)},
\end{equation}
\begin{equation}
 \label{I-2} I_\perp =\frac{8\pi}{\Phi_0}
 \left[(j_1-j_2)(\kappa_1 j_1+\kappa_2 j_2)
  +2j_1 j_2 (\kappa_1-\kappa_2)\cos^2 \frac{\phi}{2} \right]
 \frac{E_{J0}^2}{E_J(\phi)}.
\end{equation}
Note that the operator $I_s$ is diagonal for arbitrary values of
parameters $Q_0$ and $\phi$ only if the junctions are identical,
i.e., $j_1=j_2$ and $\kappa_1=\kappa_2$, so $I_\perp=0$.

The additional magnetic energy term in the system Hamiltonian is
\begin{equation} \label{Hm} H_m=
(\Phi_0/2\pi)^2 (\phi-\phi_e)^2/2L =LI_s^2(\phi,\chi)/2.
\end{equation}
The kinetic energy term associated with the charging of the chain
of the qubit junctions connected in series, $H_{\tilde{c}}=
(\Phi_0/2\pi)^2 \tilde{C}\dot{\phi}^2/2$, where
$\tilde{C}=C_{\rm{qubit}}+C_{\rm{stray}}$. Although the
capacitance between the ends of the SQUID loop $C_{\rm{stray}}$
depends on the sample layout, it clearly dominates over the first
term, $C_{\rm{qubit}}=C_1 C_2/C_{\Sigma}$ which is typically about
1\,fF \cite{Brighton}. The resonance frequency
$\omega_L=(L\tilde{C})^{1/2}$ of the oscillator formed by $L$ and
$\tilde{C}$ is therefore relatively low. Assuming $\hbar\omega_L
\sim k_B T$, but still $\omega_L \gg \omega_{\rm rf}$ we recover
classical behavior of $\phi$. In this case it is sufficient to
take into account only the magnetic term associated with the
circulating current given by Eq.\,(\ref{Hm}).

The phase variable $\phi$ is a single-valued function of $\phi_e$
for arbitrary values of $Q_0$, provided $L$ is sufficiently small,
viz.,
\begin{equation} \label{LL01}
L|L_{0,1}^{-1}| \leq L|L_{0,1}^{-1}(D)| = \frac{j_1
j_2}{|j_1-j_2|}\beta_L <1,
\end{equation}
where the screening parameter is
\begin{equation} \label{beta}
\beta_L=\frac{2\pi}{\Phi_0}L I_{c0}.
\end{equation}
One can see that for an almost symmetric qubit $(j_1 \approx j_2)$
the condition of smallness of inductance $L$, i.e. $\beta_L <
4|j_1-j_2|$, is more stringent than that for ordinary rf SQUIDs,
i.e. $\beta_L < 1$ \cite{Hansma}.

To derive the dependence $\phi(\phi_e)$, one can solve
Eq.\,(\ref{LIs}) by iteration. The first step gives the expression
\begin{equation} \label{phi} \phi \approx \phi_e
-(2\pi/\Phi_0)LI_s(\phi_e,\chi).
\end{equation}
Then the corresponding one-dimensional Schr\"{o}dinger equation,
\begin{equation} \label{SchrEq} \left\{-4E_c \left(\frac{\partial}{\partial
\chi} -i \frac{Q_0}{2e}\right)^2 - E_J(\phi_e)\cos \chi -
\frac{L}{4}\left[I_\parallel(\phi_e) \cos\chi+ I_\perp(\phi_e)\sin
\chi \right]^2 \right\} \psi = E \psi,
\end{equation}
is again of the type describing a quantum particle moving in a
periodic potential. In the case of $L=0$, this potential is
harmonic and the equation is of the Mathieu type \cite{Abr-Steg}.
In the case of nonzero $L$, the potential energy includes a
(small) second harmonic admixture. The eigenvalues and
eigenfunctions of this equation are found by numerical methods.

\begin{figure}
\begin{center}
\includegraphics[width=9cm]{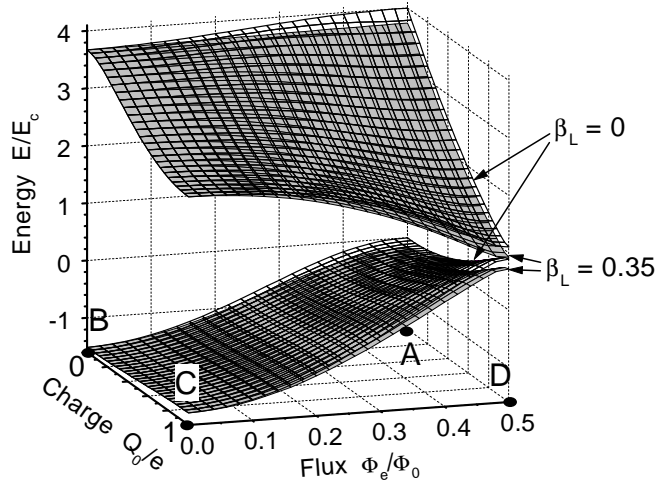}
\end{center}
\caption{Energies $E_0$ and $E_1$ calculated for the zero (wire
frames) and nonzero (surface plots) values of screening parameter
$\beta_L$ for the case of mean Josephson coupling $E_{J0}\equiv
\frac{1}{2}(E_{J1}+E_{J2})=2E_c$ and the asymmetry parameter
$a\equiv j_1-j_2=(E_{J1}-E_{J2})/(E_{J1}+E_{J2})=0.1$. Black dots
mark on the $Q_0\textrm{-}\Phi_e$ plane the locations of the magic
points $A, B,$ and $C$ and the avoided level-crossing point $D$.}
\label{E0E1}
\end{figure}

Figure\,2 shows the two lowest eigenvalues of energy (the qubit
levels) computed from the Schr\"{o}dinger equation
Eq.\,(\ref{SchrEq}). For the given parameters of the qubit and not
very small value of $\beta_L$, the subsequent terms of iteration
give substantially smaller correction. One can see that similar to
the conclusion of Ref.\,\cite{Crankshaw}, the first order effect
of inductance reduces to the lowering of both energy surfaces,
$E_0$ and $E_1$. The correction to the level spacing
Eq.\,(\ref{epsilon}) is, however, very small: $\epsilon
\rightarrow \epsilon + \Delta\epsilon_L,\: \Delta\epsilon_L \ll
\epsilon$. As can be seen from Fig.\,2, the effect of the
deformation of the energy surfaces is most appreciable in the
region of small effective Josephson coupling combined with charge
degeneracy, e.g., in the vicinity of point $D$. Around the magic
points $A$ and - especially - $B$ and $C$, which are most suitable
for qubit operation, the change of local curvature of the surfaces
is rather small.

\section{Qubit dephasing}

The term in the system Hamiltonian which describes coupling of the
qubit and the variable $\delta \Phi_e = M I_f$ associated with the
external flux noise induced in the ring by the tank circuit is
given by
\begin{equation} \label{Hcoupl} H_{\rm coupl}= - I_s\, \delta \Phi_e,
\end{equation}
or, in the qubit basis Eq.\,(\ref{epsilon}),
\begin{equation} \label{Hcoupl-q} H_{\rm coupl}=-\frac{1}{2}U_z
\sigma_z - \frac{1}{2}U_y \sigma_y = -\frac{1}{2}
\left[(c_{11}-c_{00})I_{\parallel}(\phi_e) \sigma_z + 2s_{01}
I_{\perp}(\phi_e) \sigma_y \right]\,\delta \Phi_e,
\end{equation}
where corresponding matrix elements are denoted by
\begin{equation} \label{Matr-el}
c_{00}= \langle 0|\cos \chi|0\rangle, \quad c_{11}= \langle 1|\cos
\chi|1\rangle,\quad \textrm{and} \quad s_{01} = |\langle 0|\sin
\chi|1\rangle|.
\end{equation}
The remarkable property of the coupling described by
Eq.\,(\ref{Hcoupl-q}) implies that the longitudinal term, $U_z
\propto I_{\parallel}$, vanishes in the optimal points $A$, $B$,
$C$ and $D$, i.e. where $\sin \phi_e=0$ (see Eq.\,(\ref{I-1}))
\cite{rem-Vion}. The transversal term, $U_y \propto I_{\perp}$, in
these points is, however, nonzero, and its low-frequency ($\ll
\Omega$) noise component can dephase the qubit due to
renormalization of the level spacing \cite{Makh-Shnir}, $\epsilon
\rightarrow (\epsilon^2 + U_y^2)^{1/2} \approx \epsilon +
U_y^2/2\epsilon$, yielding the diagonal term
\begin{equation} \label{X} - \frac{1}{2}X \sigma_z \equiv
- \frac{U_y^2}{2\epsilon} \sigma_z=  - \frac{2}{\epsilon}s_{01}^2
I_{\perp}^2(\phi_e)  (\delta \Phi_e)^2 \sigma_z.
\end{equation}
The problem of qubit dephasing due to quadratic longitudinal
coupling to the environment was addressed by Makhlin and Shnirman
in Ref.\,\cite{Makh-Shnir} where they focused on the
experimentally relevant $1/f$ and the Johnson-Nyquist noise power
spectra. Here we extend this model to our case of a structured
bath (low-frequency resonance tank circuit).

For white noise $I_f$ associated with losses in resistance $R_T$,
the power spectrum density $S_{\Phi}(\omega)$ of variable $\delta
\Phi_e$ is simply found from the network consideration. Neglecting
small detuning of the tank due to coupling to the qubit ring
($\omega_0 \approx \omega_1 \approx \omega_T$) we arrive at the
expression,
\begin{equation} \label{S-phi} S_{\Phi}(\omega)=
M^2 \frac{k_B T}{\pi R_T}
\frac{\omega^4_T}{(\omega^2-\omega^2_T)^2+\omega^2 \omega^2_T
Q^{-2}},\qquad -\infty< \omega <\infty,
\end{equation}
where $Q=\omega_T C_T R_T=R_T/\omega_T L_T$ is the quality factor.
Assuming that the noise is Gaussian we find the power spectrum
density of fluctuations of the variable $(\delta \Phi_e)^2$ (see,
for example, Ref.~\cite{Akhmanov}),
\begin{equation} \label{S-phi^2} S_{\Phi^2}(\omega)
=2\int^{+\infty}_{-\infty} d\omega'
S_{\Phi}(\omega')S_{\Phi}(\omega'-\omega).
\end{equation}
Taking this integral at $\omega \rightarrow 0$ yields the value
\begin{equation} \label{S_phi_0} S_{\Phi^2}(0)=
\frac{1}{\pi} M^4 Q^3 \omega_T\left(\frac{k_B T}{R_T} \right)^2,
\end{equation}
which determines the power spectrum density $S_X(0)$ of variable
$X$,
\begin{equation} \label{S_X} S_{X}(0)= \frac{2}{\epsilon^2}
s_{01}^4 I_{\perp}^4 S_{\Phi^2}(0),
\end{equation}
and the rate of "pure" dephasing in the Bloch-Redfield
approximation \cite{Shnirman}, $\Gamma_\varphi = S_X(0)/\hbar^2$.
Finally, the qubit quality factor $Q_{\varphi}\equiv
\pi\Omega/\Gamma_\varphi$ can be presented as
\begin{equation} \label{Q_phi} Q_{\varphi} =\frac{\pi}{2}
\left(\frac{I_{c0}}{s_{01}I_\perp}\right)^4
(k^2Q\beta_L)^{-2}\frac{\epsilon^3\hbar
\omega_T}{(E_{J0}k_BT)^2}\,Q,
\end{equation}
where $k=M/(LL_T)^{1/2}$ is the dimensionless coupling
coefficient.

To evaluate $Q_\varphi$ for typical parameters of the qubit
(leading for the operation points $A$, $B$ and $C$ to the value
$(\epsilon/E_{J0})^3 \sim 10$) let us assume that the quality
factor of the tank circuit is $Q \sim 300$ while the product
$k^2Q\beta_L \sim 10$ which ensures sufficient resolution in
determining the resonance frequencies $\omega_{0,1}$. Taking the
value of ratio $E_{J0}\hbar \omega_T/(k_BT)^2 \sim 0.1$ (with
$E_{J0} = 100\,\mu$eV, $\omega_T/2\pi = 100$\,MHz and $T=1$\,K) we
get an estimate: $Q_\varphi \sim 10^2
(I_{c0}/s_{01}I_{\perp})^{4}$.

In the optimal points $A$ and $D$, the amplitude $I_{\perp}$ (see
Eq.\,(\ref{I-2})) is equal to
\begin{equation} \label{Iperp-AD} (I_{\perp})_{A,D} =2
(\kappa_1 j_1+\kappa_2 j_2) I_{c0}\approx I_{c0},
\end{equation}
while in points $B$ and $C$
\begin{eqnarray} \label{Iperp-BC} (I_{\perp})_{B,C}&& =
2\left[(j_1 - j_2)(\kappa_1 j_1+\kappa_2 j_2)
+2j_1j_2(\kappa_1-\kappa_2)\right] I_{c0} \nonumber\\
&&\approx (j_1 - j_2 +\kappa_1-\kappa_2) I_{c0}.
\end{eqnarray}
The values of the matrix element $s_{01}$ in points $A$ and $B$
are equal to \cite{Zorin-JETP}
\begin{equation} \label{s01-AB} (s_{01})_{A,B} \approx
\frac{|j_1 \mp j_2|}{8\sqrt{2}}\frac{E_{J0}}{E_c},
\end{equation}
while in points $C$ and $D$
\begin{equation} \label{s01-CD} (s_{01})_{C,D} \approx 0.5.
\end{equation}

Summarizing, the order-of-magnitude estimates for parameter
$Q_\varphi$ in different operation points are
\begin{eqnarray} \label{Q-ABCD}
&&(Q_{\varphi})_{A} \sim 10^4 (j_1-j_2)^{-4}, \\
&&(Q_{\varphi})_{B} \sim 10^4 (j_1-j_2+\kappa_1-\kappa_2)^{-4}, \\
&&(Q_{\varphi})_{C} \sim 10 (j_1-j_2+\kappa_1-\kappa_2)^{-4}, \\
&&(Q_{\varphi})_{D} \sim 10.
\end{eqnarray}
For the Josephson junctions with a reasonably high symmetry in
their parameters, say $\pm 10\%$ (e.g., $j_1-j_2 = 0.1$), the
expected value of $Q_\varphi$ in points $A$, $B$ and $C$ greatly
exceeds the level $10^5$ which should be sufficient for
realization of active schemes for compensation of decoherence. In
contrast to these figures, the value achieved in the avoided
level-crossing point $D$ is relatively low.

\section{Conclusion}

We have shown that qubit dephasing due to nonlinear coupling in
the vicinity of magic operation points is relatively weak if
asymmetry of the junction parameters is sufficiently small and the
effective temperature of the tank circuit sufficiently low. Since
this temperature is associated with the backaction noise of a
preamplifier, its noise figure should be sufficiently low, as, for
example, in the case of a cold semiconductor-based amplifier or a
dc-SQUID-based amplifier \cite{Mueck}. Further improvement (if
necessary) of the charge-phase qubit with radio frequency readout
can be achieved by using a stage ensuring variable coupling
between the ring and the tank circuit. This could be, for example,
a flux transformer with a variable transfer function
\cite{Cosmelli}.

In conclusion, sufficiently symmetric parameters of the junctions
make it possible to effectively protect the charge-phase qubit
against dephasing caused by its coupling to the radio-frequency
readout circuit. The dominating mechanisms which may still cause
appreciable dephasing in this kind of qubit are most likely
related to intrinsic characteristics of the Josephson junctions
(see, e.g., Refs.\,\cite{Paladino,VanHarlingen}).

\section{Acknowledgment}

I would like to thank Alec Maassen van den Brink, Apostol Vourdas
and Denis Vion for helpful comments. This work was partially
supported by the EU through the SQUBIT-2 project.


\begin{thebibliography}{99}
\small
\setlength{\parskip}{-3.7pt}

\bibitem{Zorin-PhysC} A.\,B.~Zorin, "Cooper-pair qubit and Cooper-pair
electrometer in one device," {\it Physica C} {\bf 368}, 284-288
(2002).

\bibitem{Ilichev-FNT} E. Il'ichev, A.Yu. Smirnov, M.~Grajcar, A.~Izmalkov, D.~Born,
N.~Oukhanski, Th.~Wagner, W.~Krech, H.-G.~Meyer, and
A.M.~Zagoskin, "Radio-frequency method for investigation of
quantum properties of superconducting structures,"
cond-mat/0402559 and references therein.

\bibitem{Lupascu} A. Lupa\c{s}cu, C.\,J.\,M. Verwijs, R.\,N. Schouten, C.\,J.\,P.\,M.
Harmans, and J.\,E. Mooij, "Nondestructive readout for a
superconducting flux qubit," cond-mat/0311510.

\bibitem{Siddiqi} I. Siddiqi, R. Vijay, F. Pierre, C.M. Wilson, M. Metcalfe, C.
Rigetti, L. Frunzio, and M.H.~Devoret, "An RF-driven Josephson
bifurcation amplifier for quantum measurements," cond-mat/0312623.

\bibitem{Averin-QND} D.\,V.~Averin, "Quantum nondemolition measurements
of a qubit," {\it Phys. Rev. Lett.} {\bf 88}, 207901 (2002).

\bibitem{Bouchiat}
V.~Bouchiat, D.~Vion, P.~Joyez, D.~Esteve and M.\,H.~Devoret,
"Quantum coherence with a single Cooper pair," {\it Phys. Scr.}
{\bf T76}, 165-170 (1998).

\bibitem{Vion} D.~Vion, A.~Aassime, A.~Cottet, P.~Joyez, H.~Pothier, C.~Urbina,
D.~Esteve and M.\,H.~Devoret, "Manipulating the quantum state of
an electrical current," Science {\bf 296}, 886-889 (2002).

\bibitem{Zorin-JETP} A.\,B.~Zorin, "Josephson charge-phase qubit with radio
frequency readout: coupling and decoherence," {\it Zh. \'{E}ksp.
Teor. Fiz. } {\bf 125}(6), 1423-1435 (2004) [{\it JETP} {\bf
98}(6), 1250-1261 (2004)].

\bibitem{Krech} W.~Krech, M.~Grajcar, D.~Born, I.~Zhilyaev,
Th.~Wagner, E.~Il'ichev, and Ya.~Greenberg, "Dynamic features of a
charge qubit closed by a superconducting inductive ring," {\it
Phys. Lett. A} {\bf 303}, 352-357 (2002).

\bibitem{Rifk} R.~Rifkin and B.\,S.~Deaver, Jr., "Charge-phase relation and
phase-dependent conductance of superconducting point contacts from
rf impedance measurements," {\it Phys. Rev. B} {\bf 13}(9),
3894-3901 (1976).

\bibitem{Crankshaw} D.\,S.~Crankshaw and T.\,P.~Orlando, "Inductance
effects in the persistent current qubit," {\it IEEE Appl.
Supercond.} {\bf 11}(1), 1006-1009 (2001).

\bibitem{MaassenVanDeBrink} A. Maassen van den Brink, "Hamiltonian
for coupled qubits," cond-mat/0310425.

\bibitem{Brighton} In the rf-SQUID circuits comprising Josephson weak-links
instead of tunnel junctions a sufficiently small value of
effective capacitance $\tilde{C}$ and, hence high $\omega_L$ may
cause quantum behavior of phase $\phi$; see, for example,
R.~Whiteman, V.~Sch\"ollmann, M.~Everitt, T.\,D.~Clark,
R.\,J.~Prance, H.~Prance, J.~Diggins, G.~Buckling, and
J.\,F.~Ralf, "Adiabatic modulation of a superconducting quantum
interference device (SQUID) ring by an electromagnetic field,"
{\it J. Phys.: Condens. Matter} {\bf 10}, 9951-9968 (1998).

\bibitem{Hansma} P.\,K.~Hansma, Superconducting single-junction interferometers
with small critical currents, {\it J. Appl. Phys.} {\bf 44}(9),
4191-4194 (1973).

\bibitem{Abr-Steg} \it Handbook of Mathematical Functions, \rm edited
by M.~Abramowitz and I.\,A.~Stegun (U.S. GPO, Washingtin, D.C.,
1972), Chapter 20.

\bibitem{rem-Vion} In particular, quantum manipulation at a remarkably low
dephasing rate of the Quantronium qubit was performed in
operation point $C$ \cite{Vion}.

\bibitem{Makh-Shnir} Yu.~Makhlin and A.~Shnirman, "Dephasing of solid-state
qubits at optimal points," cond-mat/0308297.

\bibitem{Akhmanov} S.\,A.~Akhmanov, Yu.\,E.~Dyakov
and A.\,S.~Chirkin, \it Vvedenie v Statisticheskuyu Radiofiziku i
Optiku (Introduction to Statistical Radiophysics and Optics), \rm
(Nauka, Moscow, 1981), Chapter 5 - in Russian.

\bibitem{Shnirman} A.~Shnirman, Yu.~Makhlin and G.~Sch\"on, Noise and
decoherence in quantum two-level systems, {\it Phys. Scr.} {\bf
T102}, 147-154 (2002).

\bibitem{Mueck} M.~M\"uck, J.\,B.~Kycia and J.~Clarke, "Superconducting
quantum interference device as a near-quantum-limited amplifier at
0.5 GHz," {\it Appl. Phys. Lett.} {\bf 78}(7), 967-969 (2001).

\bibitem{Cosmelli} C.~Cosmelli, M.\,G.~Castellano, F.~Chiarello, R.~Leoni,
D.~Simeone, G.~Torrioli, and P.~Carelli, "Controllable flux
coupling for the integration of flux qubits," cond-mat/0403690.

\bibitem{Paladino} E.~Paladino, L.~Faoro, G.~Falci, and R.~Fazio, "Decoherence
and $1/f$ noise in Josephson qubits," {\it Phys. Rev. Lett.} {\bf
88}, 228304 (2002).

\bibitem{VanHarlingen} D.\,J.~Van Harlingen, T.\,L.~Robertson, B.\,L.\,T.~Plourde,
P.\,A.~Reichardt, T.\,A.~Crane, and J.~Clarke, "Decoherence in
Josephson-junction qubits due to critical current fluctuations,"
cond-mat/0404307.


\end{thebibliography}
\end{document}